\documentclass[]{aastex631}
\usepackage{amsmath}
\usepackage{colortbl}
\usepackage{multirow}
\usepackage{subfigure}
\usepackage{graphicx}
\newcommand{\R}{\textcolor[rgb]{1.00,0.00,0.00}}
\newcommand{\G}{\textcolor[rgb]{0.00,1.00,0.00}}
\newcommand{\uv}{$u$-$v$}
\newcommand{\klmd}{\,k$\lambda$}
\newcommand{\cm}{\,cm}
\definecolor{DarkGreen}{rgb}{0.0,0.5,0.0}
\shorttitle{Liu article}
\shortauthors{Liu et al.}
\begin{document}
\title{A Spatially resolved X-ray Polarization map of the Vela Pulsar Wind Nebula}
\author[0009-0007-8686-9012]{Kuan Liu}
\affiliation{Guangxi Key Laboratory for Relativistic Astrophysics, School of Physical Science and Technology, Guangxi University, Nanning 530004, China.}
\author[0000-0002-0105-5826]{Fei Xie}
\correspondingauthor{Fei Xie}
\email{xief@gxu.edu.cn}
\affiliation{Guangxi Key Laboratory for Relativistic Astrophysics, School of Physical Science and Technology, Guangxi University, Nanning 530004, China.}
\affiliation{INAF Istituto di Astrofisica e Planetologia Spaziali, Via del Fosso del Cavaliere 100, 00133 Roma, Italy}
\author{Yi-han Liu}
\affiliation{Department of Physics, The University of Hong Kong, Pokfulam, Hong Kong}
\affiliation{School of Physics and Astronomy, Sun Yat-Sen University, Guangzhou, 510275, China}

\author[0000-0002-5847-2612]{Chi-Yung Ng}
\affiliation{Department of Physics, The University of Hong Kong, Pokfulam, Hong Kong}
\author[0000-0002-8848-1392]{Niccol\`{o} Bucciantini}
\affiliation{INAF Osservatorio Astrofisico di Arcetri, Largo Enrico Fermi 5, 50125 Firenze, Italy}
\affiliation{Dipartimento di Fisica e Astronomia, Universit\`{a} degli Studi di Firenze, Via Sansone 1, 50019 Sesto Fiorentino (FI), Italy}
\affiliation{Istituto Nazionale di Fisica Nucleare, Sezione di Firenze, Via Sansone 1, 50019 Sesto Fiorentino (FI), Italy}
\author[0000-0001-6711-3286]{Roger W. Romani}
\affiliation{Department of Physics and Kavli Institute for Particle Astrophysics and Cosmology, Stanford University, Stanford, California 94305, USA}
\author[0000-0002-5270-4240]{Martin C. Weisskopf} 
\affiliation{NASA Marshall Space Flight Center, Huntsville, AL 35812, USA}
\author[0000-0003-4925-8523]{Enrico Costa}
\affiliation{INAF Istituto di Astrofisica e Planetologia Spaziali, Via del Fosso del Cavaliere 100, 00133 Roma, Italy}
\author[0000-0003-0331-3259]{Alessandro Di Marco} 
\affiliation{INAF Istituto di Astrofisica e Planetologia Spaziali, Via del Fosso del Cavaliere 100, 00133 Roma, Italy}
\author[0000-0001-8916-4156]{Fabio La Monaca}
\affiliation{INAF Istituto di Astrofisica e Planetologia Spaziali, Via del Fosso del Cavaliere 100, 00133 Roma, Italy}
\author[0000-0003-3331-3794]{Fabio Muleri}
\affiliation{INAF Istituto di Astrofisica e Planetologia Spaziali, Via del Fosso del Cavaliere 100, 00133 Roma, Italy}
\author[0000-0002-7781-4104]{Paolo Soffitta}
\affiliation{INAF Istituto di Astrofisica e Planetologia Spaziali, Via del Fosso del Cavaliere 100, 00133 Roma, Italy}
\author[0000-0002-9370-4079]{Wei Deng}
\affiliation{Guangxi Key Laboratory for Relativistic Astrophysics, School of Physical Science and Technology, Guangxi University, Nanning 530004, China.}
\author{Yu Meng}
\affiliation{Guangxi Key Laboratory for Relativistic Astrophysics, School of Physical Science and Technology, Guangxi University, Nanning 530004, China.}
\author[0000-0002-7044-733X]{En-wei Liang}
\affiliation{Guangxi Key Laboratory for Relativistic Astrophysics, School of Physical Science and Technology, Guangxi University, Nanning 530004, China.}

\begin{abstract} 
In this paper, we present a full spatially resolved polarization map for the Vela Pulsar Wind Nebula (PWN) observed by IXPE.
By employing effective background discrimination techniques, our results show a remarkably high degree of local polarization in the outskirt region, exceeding $60\%$ ($55\%$) with a probability of 95$\%$ ($99\%$), which approaches the upper limit predicted by the synchrotron emission mechanism. 
The high degree of polarization suggests that the turbulent magnetic energy is at most 33\% of the ordered one.
In addition, the X-ray polarization map exhibits a toroidal magnetic field pattern that is consistent with the field revealed by radio observations across the entire nebula.
This consistency reveals that the observed X-ray and radio emissions are radiated by electrons from the same magnetic field. 
Different from the Crab PWN, the consistency observed in the Vela PWN may be attributed to the interaction between the reverse shock of supernova blast wave and the PWN, which leads to a displacement between the synchrotron-cooled nebula and the fresh nebula close to the pulsar. 
These findings deepen our understanding of the structure and evolution of the Vela PWN, and the magnetohydrodynamic interaction in PWNe. 
\end{abstract} 

\section{Introduction}\label{Sec:Introduction}  

A Pulsar Wind Nebula (PWN) is an intriguing astrophysical phenomenon that provides the unique environment for studying the interaction processes between the magnetized pulsar wind and the surrounding materials.
The Vela PWN, located at a distance of 290 pc (\citealp{Dodson_Legge_2003_VelaPulsarProper}), is one of the best objects for investigating the interplay between the PWN and the reverse shock from the supernova explosion. 
The nebula is powered by the Vela pulsar (also known as PSR B0833-45) with an age of 11.4 kyr \citep{Caraveo_Luca_2001_DistanceVelaPulsar}, and resides in the north of the extended radio structure Vela X. 
The Vela X is considered as a PWN relic, and the offset between the pulsar and the Vela X's center is thought to result from the interaction between the nebula and the asymmetric reverse shock from supernova blast-wave (\citealp{Blondin_Chevalier_2001_PulsarWindNebulae}; \citealp{Slane_Lovchinsky_2018_InvestigatingStructureVela}). 

\begin{figure}[htbp]
\centering
\subfigure{\includegraphics[width=0.32\textwidth, trim={0 0cm 3cm 3cm}]{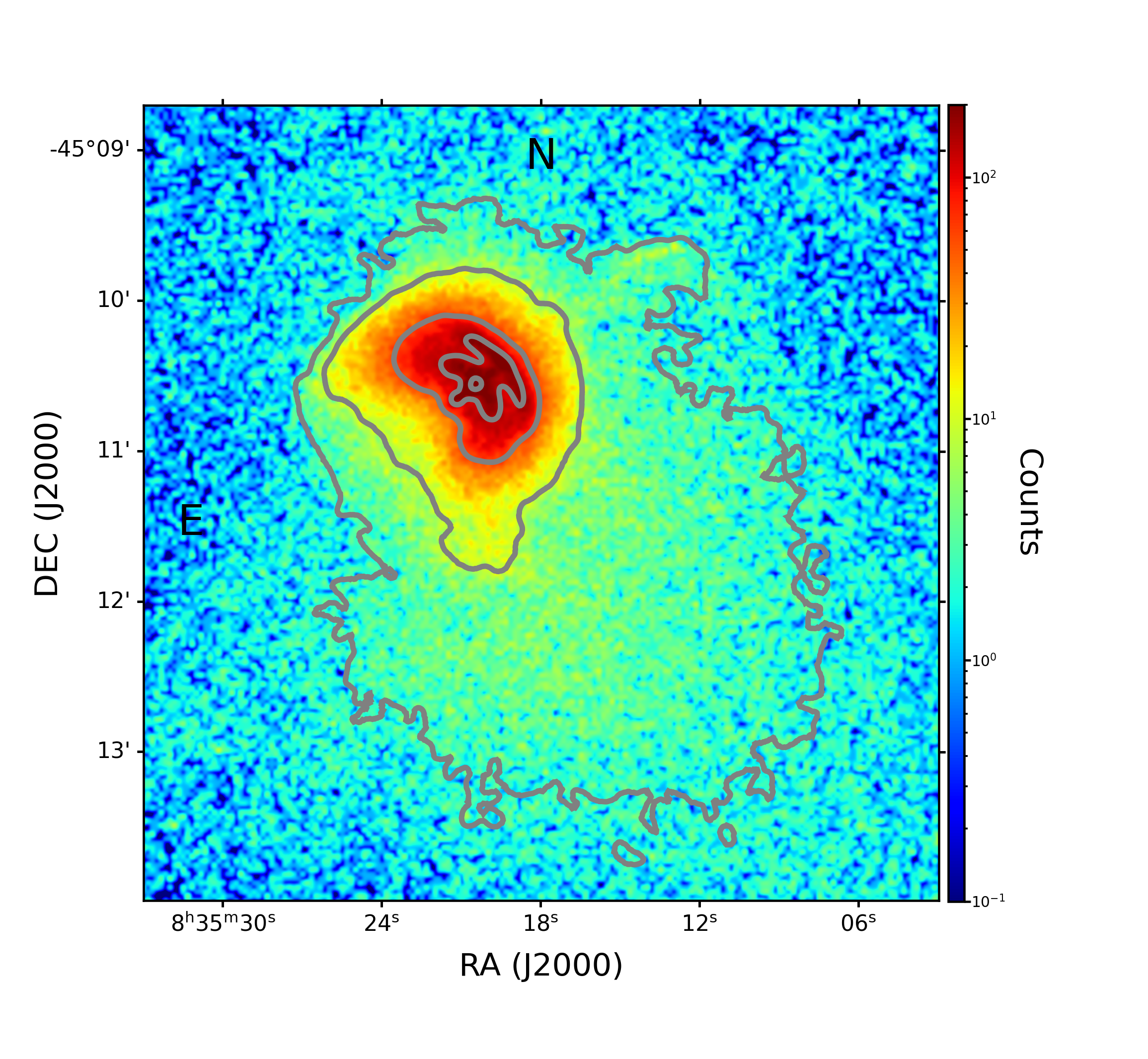}}
\subfigure{\includegraphics[width=0.335\textwidth, trim={0 1cm 3cm 0cm}, clip]{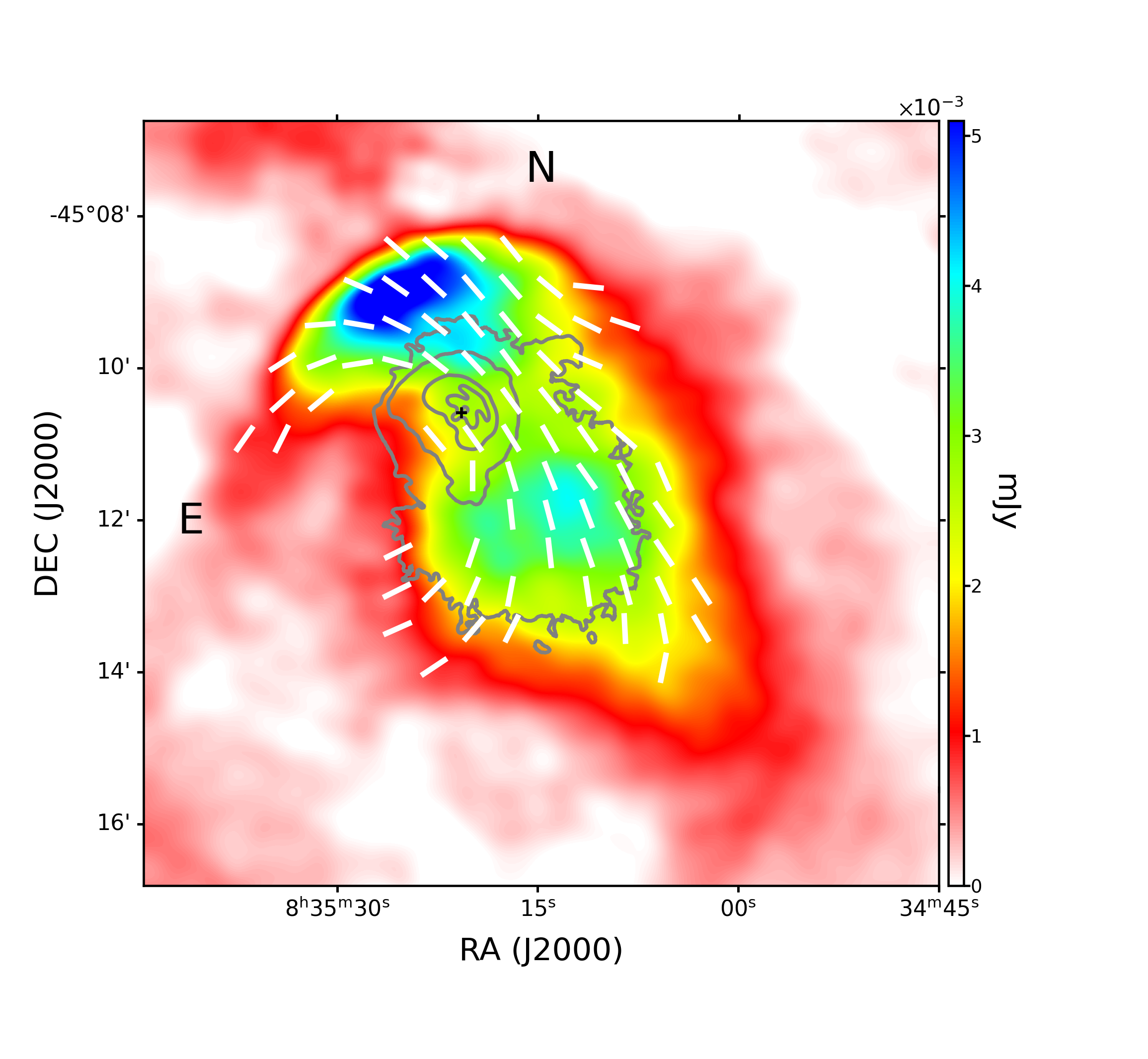}}
\subfigure{\includegraphics[width=0.32\textwidth, trim={0 0cm 3cm 0cm}, clip]{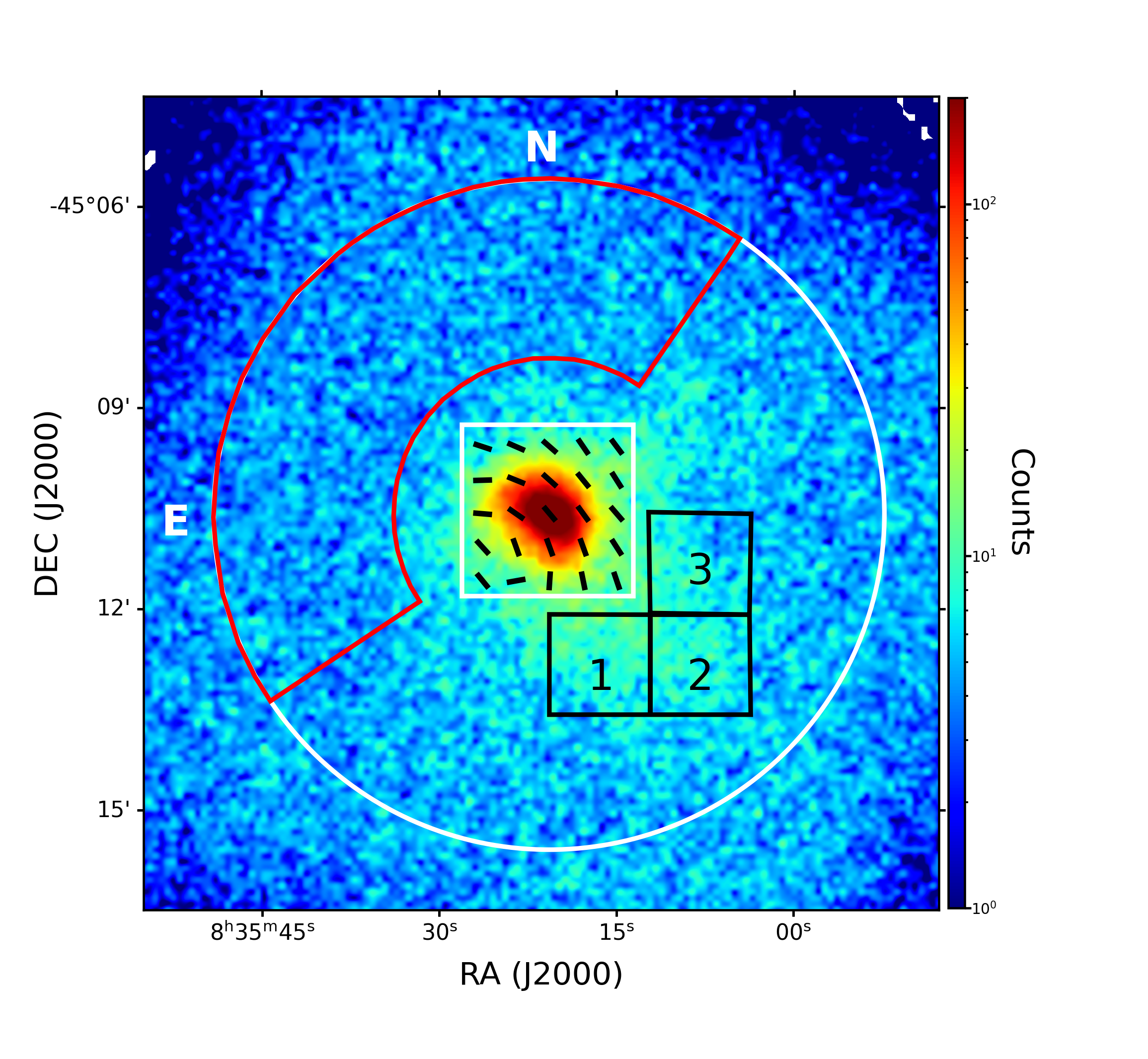}}
\caption{Chandra (left), ATCA (middle, 6 cm) and IXPE (right) images of the Vela PWN. 
In the ATCA image, the contour from the Chandra observation is overlaid. 
The white box in the right panel is the region analyzed in \cite{Xie_DiMarco_2022_VelaPulsarWind}, and the white circle marks the boundary of a radius of $300''$ centered at the pulsar. 
The red panda region and three black squared boxes identify the background and the source regions in the analysis in Section~\ref{Sec:Polarization Analysis}, respectively. 
Each black box is $90''\times90''$, and the background region extends from an inner radius of 140$''$ up to an outer radius of 300$''$ centered on the pulsar, with an open angle of 160$^\circ$.
The white (black) lines in the middle (right) panel show the magnetic field direction revealed by the radio (X-ray) polarization.
}
\label{Fig:IntroFluxMap}
\end{figure}

Multi-wavelength observations have been employed to comprehensively study and characterize the Vela PWN. 
The Chandra X-ray Observatory has provided detailed information into the innermost features (see the left panel of Fig.~\ref{Fig:IntroFluxMap}), revealing the presence of two prominent arcs and a jet toward the north-western direction within a radius of $\sim1'$ centered on the Vela pulsar (\citealp{Helfand_Gotthelf_2001_VelaPulsarIts}). 
A deeper observation shows a soft emission shell encompassing the arc-jet structure, and a faint, diffuse emission structure that extends towards the south-western region with a scale of several arc minutes (\citealp{Kargaltsev_Pavlov_2004_SpatiallyResolvedSpectrum}).
Observations in the radio band, at 5 GHz, display an extended image with two distinctive and asymmetric lobes (see the middle panel of Fig.~\ref{Fig:IntroFluxMap}).
These lobes are located in the northeast and southwest of the center pulsar, with a diameter exceeding $10^{\prime}$ (\citealp{Dodson_Lewis_2003_RadioNebulaSurrounding}). 
However, the radio nebula does not show significant features resembling the X-ray arc-jet structure observed in the central region. 
Instead, it primarily overlaps with the dim diffuse X-ray emission observed in the southwest region. 
Despite the significant differences in multi-wavelength images, the south-west region exhibits notably nebular emissions in both X-ray and radio bands. 
Thus, the presence of the south-western nebula could provide some insights into the multi-wavelength observations of Vela PWN.

Magnetic field plays a crucial role in the evolution of PWNe, which could be investigated through a polarization study.
\cite{Dodson_Lewis_2003_RadioNebulaSurrounding} showed the polarization map of the radio nebula, and a high polarization degree (PD) of 60\% was found. 
The observation revealed a large-scale toroidal magnetic field with a symmetry axis aligned with the direction of the jet axis and the pulsar proper motion. 
Recently, \cite{Xie_DiMarco_2022_VelaPulsarWind} reported a remarkably high X-ray PD in the central region (the white square region in the right panel of Fig.~\ref{Fig:IntroFluxMap}), using the Imaging X-ray Polarimeter Explorer (IXPE) (\citealp{Weisskopf_Soffitta_2022_ImagingXrayPolarimetry}; \citealp{Soffitta_Baldini_2021_InstrumentImagingXRay}). 
It is found that the X-ray PD approached the upper limit permitted by the synchrotron emission mechanism, suggesting a highly ordered magnetic field near the particle acceleration site. 
In addition, the polarization angle (PA) of the X-ray shows a good agreement with that of the radio in the central region, despite their distinct shape.  

The X-ray polarization study provides new information for understanding the formation and evolution of the Vela PWN.
In this paper, we present a full X-ray polarization study for the entire Vela PWN.
We extend the findings presented in \cite{Xie_DiMarco_2022_VelaPulsarWind} by analysing the polarization characteristics observed in the south-western diffuse emission region, estimating the upper limit of the magnetic field strength ratio between the random and the ordered magnetic field, and conducting a comprehensive comparison between the X-ray and radio polarization pattern. 

\section{Observation and Data Reduction}\label{Sec:Observation and Data Reduction}

\subsection{X-ray Polarization Data}\label{SubSec:Prescription}

IXPE, a SMall EXplorers (SMEX) Mission funded by NASA and the Agenzia Spaziale Italiana (ASI), was launched on 9th December 2021 (\citealp{Weisskopf_Soffitta_2022_ImagingXrayPolarimetry}). 
It is the first spatially resolved observatory fully dedicated to X-ray polarimetry in the energy band 2--8 keV. 
IXPE consists of three identical telescopes, each one having a polarimeter, the Gas Pixel Detector (GPD) (\citealp{Costa_Soffitta_2001_EfficientPhotoelectricXray};  \citealp{Baldini_Barbanera_2021_DesignConstructionTest}), sensitive to linear X-ray polarization and placed at the focus of the co-aligned Wolter-1 mirror module assemblies (MMAs). The IXPE MMAs have an angular resolution in half power diameter (HPD) of $\sim 30''$, and a field of view (FoV) of $12.9^{\prime}\times12.9^{\prime}$.

The Vela PWN was observed on April 2022 for a total exposure of $\sim$860~ks. 
Data is publically available on NASA'S HEASARC archive, and analysis is performed with \textsc{ixpeobssim} V30.5.0 (\citealp{Baldini_Bucciantini_2022_IxpeobssimSimulationAnalysis}), which is developed by the IXPE collaboration following the formalism in \cite{Kislat_Clark_2015_AnalyzingDataXRay}.
Details on data extraction are reported in \cite{Xie_DiMarco_2022_VelaPulsarWind}.

The right panel of Fig.~\ref{Fig:IntroFluxMap} shows the images of the Vela PWN observed by IXPE, as well as the $150''\times150''$ squared region analysed in \cite{Xie_DiMarco_2022_VelaPulsarWind}, and three $90''\times90''$ black boxes in the diffuse emission region.
All regions, including source and background, are chosen within a radius of $300''$ centered on the pulsar (the white circle) to avoid the geometrical edge effects, as discussed in \cite{DiMarco_Soffitta_2023_HandlingBackgroundIXPE}.
The energy range was narrowed down to 2–5 keV for further increment of the signal to background ratio (S/N ratio, hereafter).
In addition, we applied the data selections provided in \cite{DiMarco_Soffitta_2023_HandlingBackgroundIXPE} and \cite{Xie_Ferrazzoli_2021_StudyBackgroundIXPE}, which successfully remove approximately $40\%$ of the instrumental background, reducing the background contamination significantly.

\subsection{Radio Polarization Data}

We produced also radio polarization images of Vela PWN at 6, 13, and 21\,cm for a comparison with the X-ray results. 
All of the radio data were taken from the Australia Telescope Compact Array (ATCA), and Tab.~\ref{tab:Radio_Observations_Table_Vela} shows detailed information about selected observations in every band. 
There are five observations in 6\cm\ band with the integration times of 53.7\,hr, covering a \uv~range from 0.35 to 115\klmd;
four 13\cm\ mosaic observations give 0.8\,hr integration times and a \uv\ coverage from 0.25 to 40.5\klmd;
five other observations have 5.6\,hr integrated observation time at 21\cm, which covers a \uv\ coverage from 0.1 to 29.6\klmd.

\begin{table*}[ht!]
\begin{tabular}{cccccccc}
\toprule
Telescope & Obs. Date&Array   &Center Freq.&Usable Band-&No. of   &{Integration}& Pulsar Bin- \\
 &      & Config. & (MHz) & width (MHz) & Channels & Time (hr)& ning Mode\\      
\hline
  \textbf{6\,cm}\\
\hline
ATCA & 2018 Jan 09&6C&5997.5&2048&513&9.5&Y\\
ATCA & 2001 Feb 24&375&4800, 5696&104&13&11.5&Y\\
ATCA & 2001 Mar 17&1.5D&4800, 5696&104&13&10.4&Y\\
ATCA & 2001 Mar 30&6E&4800, 5696&104&13&10.7&Y\\
ATCA & 2001 Apr 18&750D&4800, 5696&104&13&10.6&Y\\
\hline 
  \textbf{13\,cm}\\
\hline
ATCA & 1996 Jan 09&750C&2368&104&13&0.2&Y\\
ATCA & 1996 Jan 24&750B&2368&104&13&0.2&Y\\
ATCA & 1996 May 21&750D&2368&104&13&0.2&Y\\
ATCA & 1996 Nov 22&750A&2368&104&13&0.2&Y\\
\hline 
  \textbf{21\,cm}\\
\hline
ATCA & 1996 Jan 09&750C&1344&104&13&0.2&Y\\
ATCA & 1996 Jan 24&750B&1344&104&13&0.2&Y\\
ATCA & 1996 May 21&750D&1344&104&13&0.2&Y\\
ATCA & 1996 Jul 31&6C&1344, 1432&104&13&4.8&Y\\
ATCA & 1996 Nov 22&750A&1344&104&13&0.2&Y\\
\toprule
\end{tabular}
\caption{Radio observations of Vela PWN used for the polarization analysis}
\label{tab:Radio_Observations_Table_Vela}
\end{table*}

The MIRIAD package helps to reduce all the radio data in this study \citep{sault1995miriad}. 
According to the standard procedures, we flagged bad data affected by radio frequency interference, obtained calibrated observation solutions (e.g., flux, band pass, and gains) from primary and secondary calibrators, and applied these solutions to the target source for every observation. 
We also eliminated contamination from the strong pulsar flux by excluding emission in the on-pulse phase with the help of ATCA pulsar binning mode (see Tab.~\ref{tab:Radio_Observations_Table_Vela} the last column).
As is shown in Tab.~\ref{tab:Radio_Observations_Table_Vela}, every observation in an individual band was combined together to produce Stokes \textit{I}, \textit{Q}, and \textit{U} radio images of the Vela PWN. 
The rms noises of images are around 0.03, 0.04, and 0.08 \,mJy\,beam$^{-1}$ at 6, 13, and 21\cm, respectively. 
We weighted data inversely proportional to the noise level and the Brigg's robust parameter of 0.5 \citep{briggs1995american}, and plotted images in every band with beam sizes of 30\arcsec. 
Then, we deconvolved the initial images and combined final Stokes \textit{I}, \textit{Q}, and \textit{U} images to produce the PA) images of polarized emission from Vela PWN in each band. 
Finally, we linearly fit the PA of every pixel in each band to measure the Faraday effect and derive the corrected PA map of magnetic field in the radio PWN.

\section{Polarization Result}\label{Sec:Polarization Analysis}

\begin{figure}
\centering
\includegraphics[width=0.65\textwidth, trim={0 1cm 0cm 1cm}, clip]{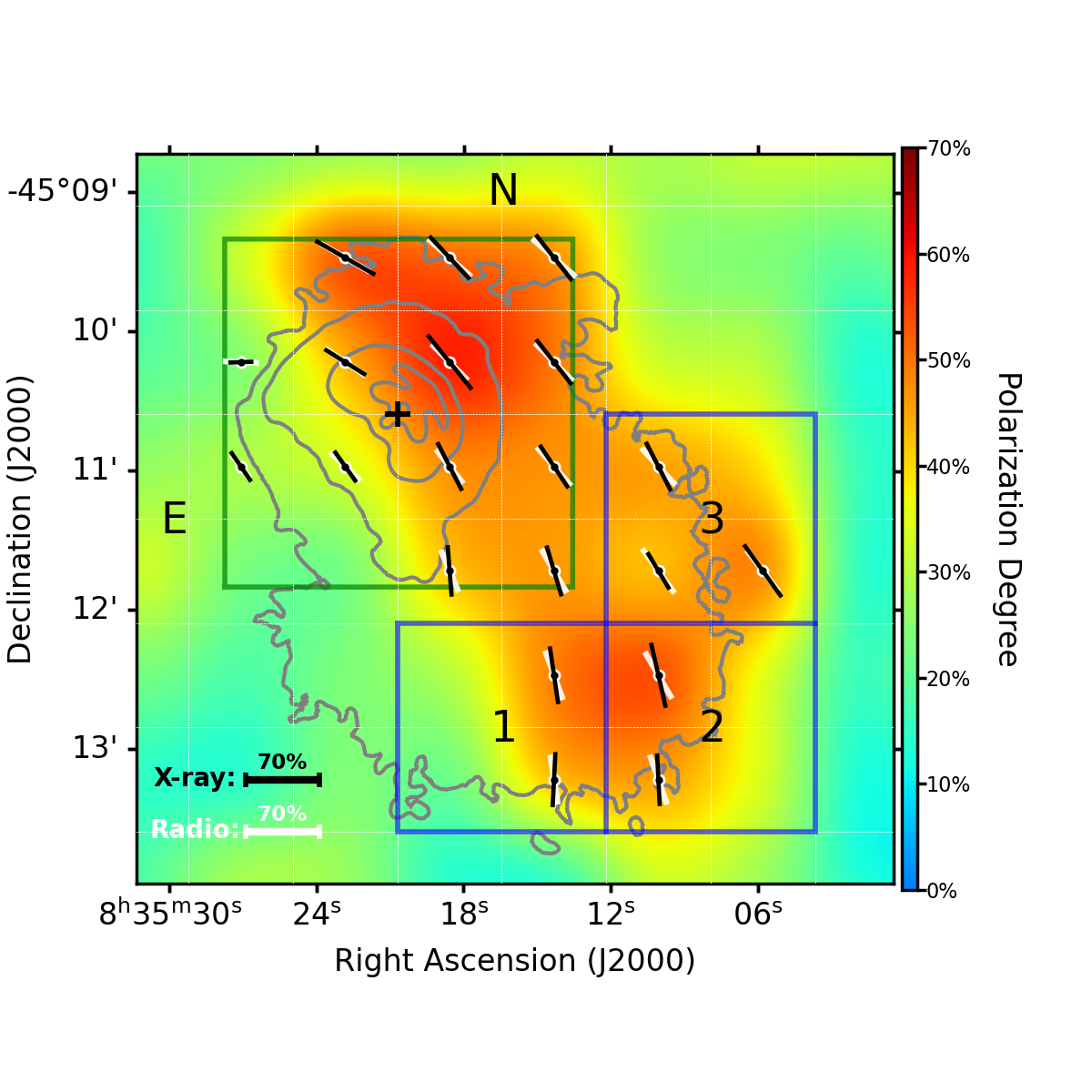} \\
\includegraphics[scale=0.08, trim={3cm 0cm -7cm 0}, clip]{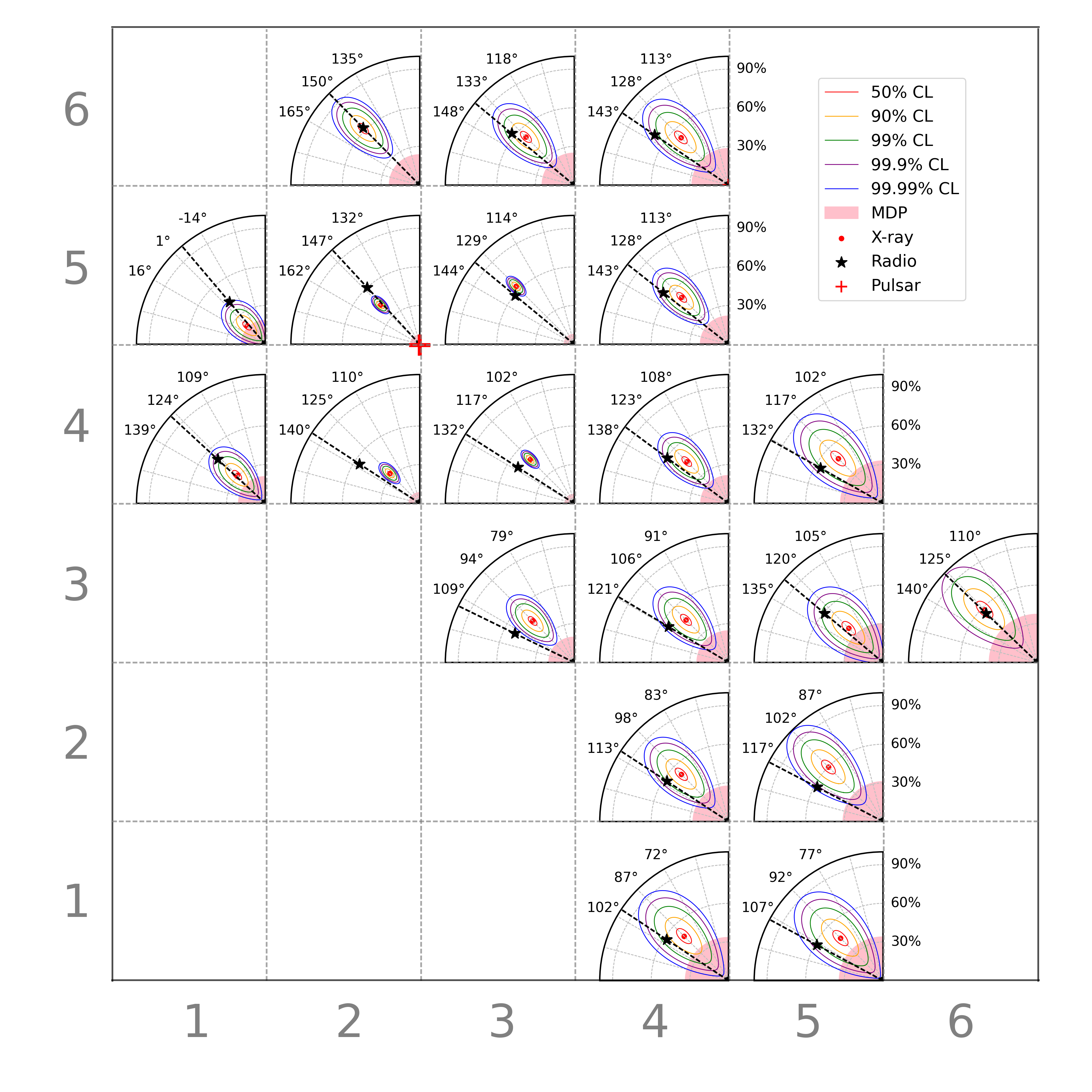}
\caption{(Top) The PD map of the Vela PWN in the 2--5\,keV energy range with a pixel size of $45''$ overlaid with the Chandra contour.
Pixels with X-ray PD $>$ MDP$_{99}$ have polarization vectors on top.
The length of the black (white) lines indicate the PD measure in X-ray (radio), and orientation indicate the projected magnetic field.
(Bottom) The polarization protractor plot for pixels having the polarization vector overlaid on the top panel, covering the same region, with radius and polar angle indicating the PD and PA, respectively.}
\label{Fig:45X45}
\end{figure}

We performed a spatially resolved polarization analysis of Vela PWN in X-ray and radio band.
The top panel of Fig.~\ref{Fig:45X45} displays the X-ray PD map using $45''\times45''$ grids, convoluted with a Gaussian kernel function with a kernel size of $67.5''$, and overlaid with Chandra contours in the 2-5 keV band. 
The green box marks the region analysed in \cite{Xie_DiMarco_2022_VelaPulsarWind}, and three blue boxes are the outskirt regions we focus on (same as the 1, 2, 3 labeled square regions in Fig~\ref{Fig:IntroFluxMap}). 
The direction of the black lines represent the projected magnetic field, perpendicular to the X-ray PA (defined from north to east), with its length proportional to the PD.
The white lines indicate the Faraday-rotation corrected magnetic field derived from radio data \citep{liu_radio_2023}.
Only the grids with X-ray PD larger than MDP$_{99}$ (i.e., the minimum detectable polarization at 99\% confidence level) have polarization vectors overlaid, and the corresponding polarization protractor plots are presented in the bottom panel.
In these protractor plot, the confidence levels of 50\%, 90\%, 99\%, 99.9\%, and 99.99\% are enclosed by the red, yellow, green, purple and blue contours, and the MDP$_{99}$ are filled with pink shadow.
The radio polarization is marked by the black star, and the black dash line highlight the direction of the radio PA. 
Owing to that the error of radio PA is at the level of 1$^\circ$, which is much smaller than that of X-ray, it is neglected in our discussion. 

\cite{Xie_DiMarco_2022_VelaPulsarWind} has shown that the X-ray polarization is well consistent with the radio in the central region.
In this paper, we extend this consistance to the entire nebula. 
The maximum discrepancy in PA between the X-ray and the radio is below $20^\circ$, and the average PA difference is $\sim 17^\circ$ with a mean error of $5^\circ$. 
In the areas exhibiting the most difference, such as pixels (5, 2) and (5, 1), the X-ray PA errors are notably large, implying that the discrepancies may be heavily affected by a statistic effect. 
In addition, in pixel like (3, 4), the radio polarization deviates from the X-ray polarization by more than 4$\sigma$ (i.e., beyond the contour of 99.99\% confidence level). 
This region is close to the pulsar, thus it may be affected by the pulsar's polarization.
Furthermore, the geometry structure near the pulsar seen in the Chandra image are highly curved, therefore, the underlying PA may vary rapidly, leading to a discrepancy in polarization  between the radio and X-ray.
Future X-ray polarimetry with smaller PSF will offer more insight into this matter.

The outskirt of the Vela PWN is highly polarized, as shown in Fig.~\ref{Fig:45X45},  although it is much dimer than the central region. 
To obtain statistically significant polarization results, we rebin the diffuse emission into three $90''\times90''$ squared regions, labeled by 1, 2, and 3, and each of them contains four pixels. 
All these three boxes display high level of significance exceeding 7$\sigma$ (calculated as the PD divided by its 1$\sigma$ error), which indicate a strong polarization in the diffuse emission region.
We also consider the possible systematic effect on the diffuse emission, including the PSF and polarization leakage \citep{Bucciantini_DiLalla_2023_PolarisationLeakageDuea}, using the simulation tools in ixpeobssim \citep{Baldini_Bucciantini_2022_IxpeobssimSimulationAnalysis} and GEANT4 \citep{Agostinelli_Allison_2003_Geant4SimulationToolkit}.
It is found that these effects cause negligible impact on the X-ray polarization results within 1$\sigma$ uncertainty, thus are neglected in our analysis. 

Moreover, for the area with low S/N ratio, such as the diffuse emission region, background subtraction could be crucial. 
We consider a local background sampled from the north-east side of the Vela PWN (the red panda region in the right panel of Fig.~\ref{Fig:IntroFluxMap}). 
This background is unpolarized, with an upper limit of 7.6\%.
Its normalized Stokes parameters are subtracted from the source data, with results presented in Tab.~\ref{Tab:local_extra}.
In general, the PD substantially increase to almost twice its original value after background subtraction, ranging from $\sim50\%$ to $\sim80\%$, with $1\sigma$ error of $\sim10\%$ and significance level $\sim7\sigma$. 
To enhance the significance, we combine region 2 and 3, which are two most polarized area in the diffuse emission. 
The combined region have a PD of $76.2\pm7.2\%$ with a significance level larger than $10\sigma$. 

\begin{table}
\centering
\begin{tabular}{c|ccccc}
\toprule
Region & \textit{Q}/\textit{I} & \textit{U}/\textit{I} & Sig($\sigma$) & PD($\%$) & PA($^\circ$) \\
 \hline
 1 & $-0.35\pm0.05$ & $0.02\pm0.05$ & 7.00 & $35.0\pm5.0$ & $88.1\pm4.1$ \\
 2 & $-0.37\pm0.06$ & $-0.20\pm0.06$ & 7.33 & $41.6\pm5.7$ & $104.1\pm3.9$ \\
 3 & $-0.24\pm0.06$ & $-0.38\pm0.06$ & 7.82 & $44.5\pm5.7$ & $118.8\pm3.7$ \\
Background & $0.06\pm0.02$ & $-0.03\pm0.02$ & 2.72 & $<7.6$ & $-$ \\
\hline
 \multicolumn{6}{c}{\textbf{Background Subtracted Polarization}}\\
\hline
$1$& $-0.56\pm0.08$ & $0.05\pm0.08$ & 7.21 & $56.4\pm7.8$ & $87.5\pm4.0$ \\
$2$& $-0.69\pm0.10$ & $-0.33\pm0.10$ & 7.36 & $76.7\pm10.4$ & $102.6\pm3.9$ \\
$3$& $-0.48\pm0.11$ & $-0.65\pm0.11$ & 7.60 & $80.7\pm10.6$ & $116.9\pm3.8$ \\
$2+3$& $-0.59\pm0.07$ & $-0.49\pm0.07$ & 10.49 & $76.2\pm7.2$ & $108.8\pm2.8$ \\
\toprule
\end{tabular}
\caption{The normalised stokes parameters \textit{Q}/\textit{I}, \textit{U}/\textit{I} and the significance level of the diffuse emission and background are reported, along with the background subtracted polarization.
Additionally, the PD, PA, and their uncertainties are calculated at $68.3\%$ confidence level. 
For PD $<$ MDP$_{\rm 99}$, the upper limit in terms of MDP$_{\rm 99}$ are reported.}
\label{Tab:local_extra}
\end{table}


\section{Discussions}\label{Sec:Discussions}

We detect a high PD in the outskirt region of the Vela PWN.
For the most polarized region, at a confidence level of $90\%$ ($95\%$), the PD is larger than $\sim60\%$ ($\sim55\%$) after background subtraction. 
The maximum PD allowed by the synchrotron mechanism for a power-law photon spectrum, with a photon index of $\Gamma$, is given by $\Pi=3\Gamma/(3\Gamma+2)$ (\citealp{Rybicki_Lightman_2004_RadiativeProcessesAstrophysics}). 
For the diffuse emission, with a spectral index of $1.3\sim1.4$ (\citealp{Kargaltsev_Pavlov_2004_SpatiallyResolvedSpectrum}, \citealp{Kargaltsev_Pavlov_2008_PulsarWindNebulae}, i.e., $\Gamma$ is $0.3\sim0.4$), the maximum PD is $66\sim67\%$. 
Thus the observed PD of the outer diffuse emission is close to the maximum. 
The diffuse emission is located in $2^{\prime}\sim4^{\prime}$ away from the pulsar, which is almost 10 times further than the arc structure in the center PWN (\citealp{Ng_Romani_2004_FittingPulsarWind}). 
The high PD of the diffuse emission reveals that the magnetic field in the outflow remains undisturbed as the magnetized plasma travels from the acceleration site to the outer region. 

\begin{figure}
\centering
\includegraphics[width=0.45\textwidth]{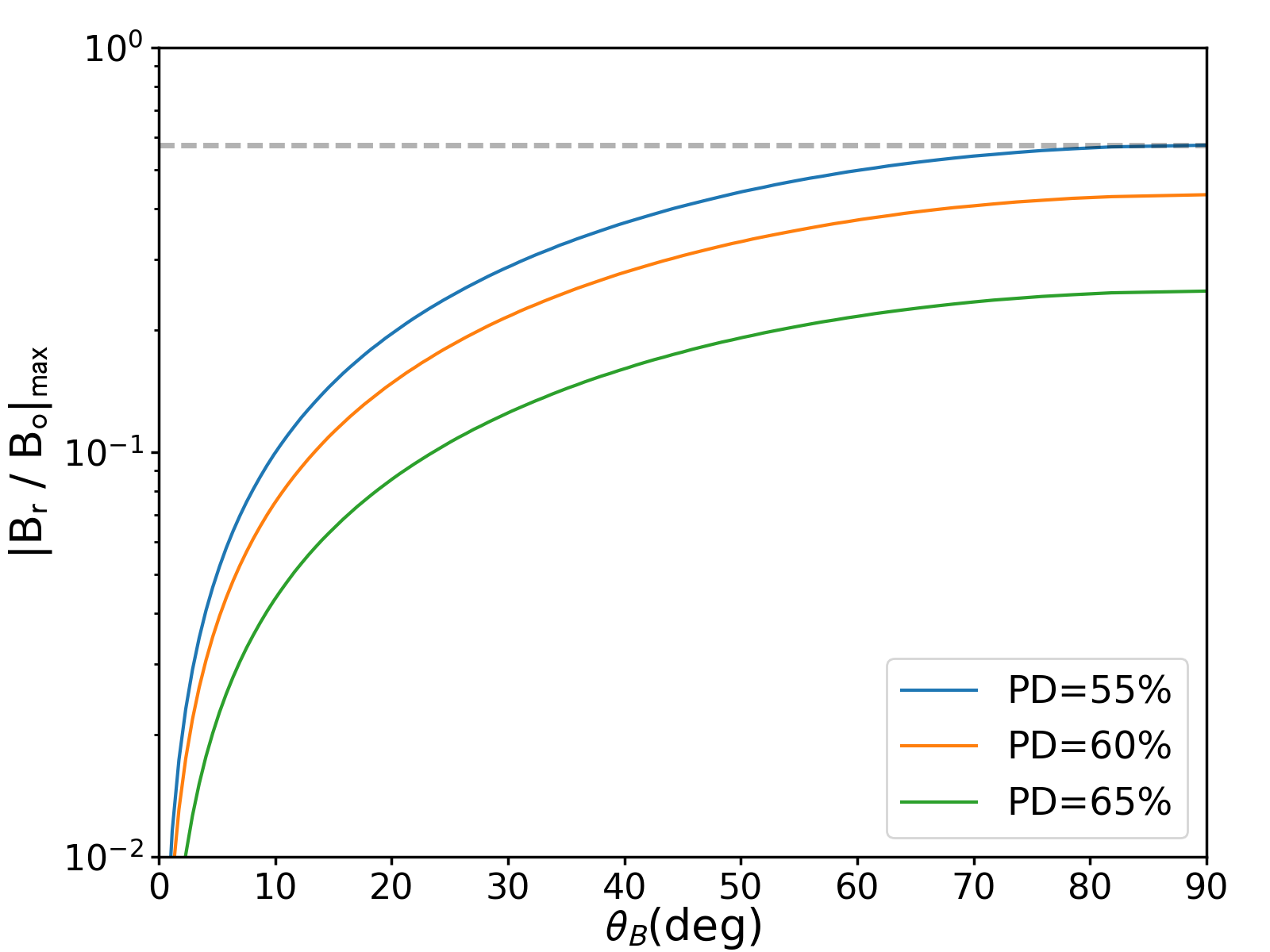}
\caption{The relationship between the upper limit of the ratio between the random and ordered magnetic field strength and the angle between the light of sight and the magnetic field direction, with given PD of $55\%$ (blue), $60\%$ (orange), and $65\%$ (green).}
\label{Fig:Turbulent_Energy}
\end{figure}

Following \cite{Bandiera_Petruk_2016_RadioPolarizationMaps} and \cite{Bucciantini_Bandiera_2017_ModelingEffectSmallscale}, we estimate the upper limit of the energy ratio between the random and ordered magnetic field.
The observed PD for a local region can be defined as: 
\begin{equation}
    \Pi=\frac{\alpha+1}{\alpha+5/3}\frac{3+\alpha}{4}\frac{\sin^2{\theta_B}}{2\sigma^2}
    \frac{{}_{1}F_{1}\left((1-\alpha)/2, 3;-\sin^2{\theta_B}/2\sigma^{2}\right)}
    {{}_{1}F_{1}\left(-(1+\alpha)/2,1;-\sin^2{\theta_B}/2\sigma^{2}\right)} 
\end{equation}
Here, $\alpha$ is the spectral index, 
${ }_{1}F_{1}(a,b;x)$ is the Kummer confluent hypergeometric function, with $a$ and $b$ representing the upper and lower parameters, and $x$ being the argument of the function,
$\theta_B$ is the angle between the magnetic field direction and line of sight in the co-moving coordinate;
$\sigma=\sqrt{E_{\rm r}/3E_{\rm o}}=B_{\rm r}/\sqrt{3}B_{\rm o}$ is related to the energy ratio between the random magnetic field and the ordered magnetic field, assuming an three dimensional isotropic Gaussian random field with a variance of $(B\sigma)^2$ in each direction;
the $B_{\rm r}$, $B_{\rm o}$, and $B$ are the strength of the random, ordered, and total magnetic field in the co-moving coordinate, respectively. 
The spectral index $\alpha$ of Vela diffuse emission is measured as $1.4$ (\citealp{Kargaltsev_Pavlov_2004_SpatiallyResolvedSpectrum}; \citealp{Kargaltsev_Pavlov_2008_PulsarWindNebulae}).
The relation between the $\theta'_B$ and the upper limit of $B_{\rm r}/B_{\rm o}$ is showed in Fig.~\ref{Fig:Turbulent_Energy}, in cases of given PD of 55\%, 60\%, and 65\%. 
With a PD of 55\% and under an extreme condition $\theta_B\sim90^\circ$, the maximum of $B_{\rm r}/B_{\rm o}$ is 57\%, corresponding to a magnetic energy ratio $E_{\rm r}/E_{\rm o}$ of 33\%. 
Considering a more possible case with higher PD and smaller $\theta_B$ (for example, PD$\sim60\%$ and $\theta_B\sim50^\circ$), the upper limit of $B_{\rm r}/B_{\rm o}$ ($E_{\rm r}/E_{\rm o}$) would be $\sim33\%$ ($\sim10\%$).

Beside the strong PD in X-ray, we also found a strong correlation between the polarization of the X-ray and the radio throughout the entire nebula, as shown in Fig.~\ref{Fig:45X45}.
Both X-ray and radio observations exhibit regions of high polarization, approaching the upper limit allowed by the synchrotron mechanism.
This supports that the X-ray and radio emissions are produced by the electrons in the same magnetic field \citep{Kargaltsev_Pavlov_2004_SpatiallyResolvedSpectrum}. 

It is worth noting that such a correlation is not observed in Crab PWN \citep{Bucciantini_Ferrazzoli_2023_SimultaneousSpacePhase, Aumont_Conversi_2010_MeasurementCrabNebula}. 
This disparity might be attributed to their different evolutionary stages. 
The Crab is a relatively young nebula at an age of $\sim$1 kyr, while the Vela is $\sim$11.4 kyr, whose reverse shock of the supernova blast wave is interacting with the nebula. 
The outer nebular bubble undergoes compression, detaching from the pulsar, and forming the PWN relic, Vela X.
In the absence of the outer nebula, we can directly observe the X-ray and radio nebula near the termination site. 
The predicted X-ray and radio polarization should be consistent, according to the magnetic hydrodynamic numerical simulations \citep{Lyubarsky__2002_StructureInnerCrab, Komissarov_Lyubarsky_2004_SynchrotronNebulaeCreated,
DelZanna_Amato_2004_AxiallySymmetricRelativistic,
Bucciantini_DelZanna_2005_PolarizationInnerRegion}.
While the young nebula, as the Crab, is expanding in its supernova ejecta and has not yet been interrupted by the reverse shock \citep{Gaensler_Slane_2006_EvolutionStructurePulsar}. 
The fresh X-ray emitting nebula is surrounded by the outer cool nebula, which predominantly radiates in the radio band, leading to the observed disparity between the radio and X-ray polarization. 

Furthermore, the high PD observed in the Vela PWN implies a limited turbulence in the magnetic field, indicating some suppression of instability. 
A strong magnetic field could play a crucial role in suppressing the growth of instability \citep{Bucciantini_Amato_2004_MagneticRayleighTaylorInstability}. 
The magnetic field in the Vela could be strengthened due to the compression of the reverse shock, resulting the observed high PD.  

The boomerang PWN G106.6+2.96 is similar with Vela PWN \citep{Kothes_Reich_2006_BoomerangPWNG106}, where the original nebula has been crushed by the reverse shock.  
Its radio observations reveal a thick toroidal polarization pattern, akin to that observed in the Vela.
If our understanding holds true, its X-ray polarization should also align closely with the radio polarization. 
Future observations will validate our conjecture. 

\section{Conclusion}\label{Sec:Conclusion}
X-ray polarization is important for understanding the magnetic fields of PWNe.  
Using data from IXPE we made the first full map of the X-ray polarization for Vela PWN as a function of position. 
In addition to the polarization detected in the inner regions closest to the pulsar, we find after careful background subtraction, statistically significant high PD is detected in the extended X-ray emission, which is close to the upper limit allowed by the synchrotron emission mechanism.
The result suggests that the magnetic field in these regions remains highly uniform, and the turbulent magnetic field strength is estimated to be up to around half of the ordered magnetic field strength in the most polarized region.

We also found a strong correlation between the X-ray and radio polarization, which implies that the X-ray and radio are emitted by the electrons in the same magnetic field.
Such consistence observed in the Vela PWN could be attributed to the interaction between the nebula and the reverse shock of the supernova blast-wave, which leads to a displacement between the synchrotron-cooled nebula and the fresh nebula close to the pulsar.
The compression by the reverse shock amplifies the magnetic field, suppresses the growth of instability, and leads to an ordered magnetic field thus the high degree of polarization. 
Future X-ray polarization observations to other similar nebular systems, like boomerang PWN G106.6+2.9, will gain a deeper understanding on it. 

\begin{acknowledgments}

We thank the anonymous referee of this work for useful comments and suggestions that improved the paper.
This work is supported by National Natural Science Foundation of China (Grant No. 12373041 and Grant No. 12133003).  
N.B. was supported by  the INAF MiniGrant  ``PWNnumpol - Numerical Studies of Pulsar Wind Nebulae in The Light of IXPE''.
C.-Y. Ng is supported by a GRF grant of the Hong Kong Government under HKU 17305419.

The Imaging X-ray Polarimetry Explorer (IXPE) is a joint US and Italian mission. 
The US contribution is supported by the National Aeronautics and Space Administration (NASA) and led and managed by its Marshall Space Flight Center (MSFC), with industry partner Ball Aerospace (contract NNM15AA18C).
The Italian contribution is supported by the Italian Space Agency (Agenzia Spaziale Italiana, ASI) through contract ASI-OHBI2017-12-I.0, agreements ASI-INAF-2017-12-H0 and ASI-INFN2017.13-H0, and its Space Science Data Center (SSDC), and by the Istituto Nazionale di Astrofisica (INAF) and the Istituto Nazionale di Fisica Nucleare (INFN) in Italy. 
This research used data products provided by the IXPE Team (MSFC, SSDC, INAF, and INFN) and distributed with additional software tools by the High-Energy Astrophysics Science Archive Research Center (HEASARC), at NASA Goddard Space Flight Center (GSFC). 
The X-ray data for the Chandra observations were downloaded from the public access site, the Chandra Data Archive (CDA). This is part of the Chandra X-ray Observatory Science Centre (CXC) which is operated for NASA by the Smithsonian Astrophysical Observatory.
The Australia Telescope Compact Array is part of the Australia Telescope, funded by the Commonwealth of Australia for operation as a National Facility, and managed by CSIRO. 

\end{acknowledgments}

\clearpage


\end{document}